\documentstyle[preprint,aps,prd]{revtex}
\begin{document}
\tightenlines
\title{Dualization of non-Abelian  $B\wedge F$  model \footnote{Accepted for
publication in {\bf Phys.Lett. B}}}

\author{A.Smailagic \footnote{E-mail address: a.smailagic@etfos.hr} }
\address{Department of Physics, 
Faculty of Electrical  Engineering \\
University of Osijek, Croatia }
\author{E. Spallucci\footnote{E-mail address:spallucci@trieste.infn.it}}
\address{Dipartimento di Fisica Teorica\break
Universit\`a di Trieste,\\
INFN, Sezione di Trieste}
\maketitle

\begin{abstract}
 We show that dualization of $B\wedge F$ models to St\"uckelberg--like 
 massive gauge theories allows a non-Abelian extension. 
 We obtain local Lagrangians 
 which are straightforward  extensions of the Abelian results.
\end{abstract}

\newpage

Recently, we have described a  procedure for dualizing Abelian
$B\wedge F$ models of arbitrary $p$-forms to a gauge invariant, massive, 
St\"uckelberg-like theories \cite{top}.
It has been shown that such dualization is a special subclass of general
$p$--dualization formulation for {\it interacting} theories. One may naturally 
ask whether these models allow a non--Abelian generalization. Earlier attempts 
to dualize non--Abelian theories have  been made within the
framework of 3D-topological models\cite{rocek} and  non-linear sigma models with 
non--Abelian isometries \cite{pkt}, \cite{quevedo}. In the former case, the 
success was partial  being subject to some fine-tuning and  in both cases
 the resulting dual Lagrangians  do not correspond to their
 Abelian versions. On the other hand, the interest in  St\"uckelberg-like gauge
 models and $B\wedge F$ non-Abelian models stems form the fact that both
 models have been independently used  as alternatives to 
 the Higgs Mechanism for vector mass generation\cite{love}, \cite{lah},
 \cite{khoud}, \cite{alm}. \\
In this note we shall investigate dualization of a non-Abelian extension of the
$B\wedge F$ model, $B$ being a Kalb-Ramond two-index tensor field and $F$ 
being field strength of a non-Abelian vector field.  With this goal in mind, 
and to set the proper path, we briefly review the dualization of the 
$B\wedge F$ model in the  Abelian case. \\

The main ingredient of the dualization procedure is a {\it parent} 
Lagrangian \cite{top} defined as

\begin{eqnarray}
L_P=&-&{1\over 2\cdot 3!} H _{\mu\nu\rho}\, H ^{\mu\nu\rho }
-{1\over 3!}\epsilon^{\lambda\mu\nu\rho} C\,
\partial_{[\,\lambda} H_{\mu\nu\rho\,] }\, 
 \nonumber\\
  &+& {m\over 3!}\, \epsilon^{\lambda\mu\nu\rho}\, \left(\,
  V_\lambda-\partial_\lambda\,\theta \,\right)\, H_{\mu\nu\rho}
  -{1\over 4}\, F_{\mu\nu}(V)\, F^{\mu\nu}(V)  
  \label{parent}
\end{eqnarray}

$V_\lambda$ is  a {\it spectator} gauge field which is inert under 
dualization.  The St\"uckeleberg field  $\theta$ has been introduced 
 to provide gauge invariance of the parent Lagrangian under transformations
 $\delta V_\mu=\partial_\mu \lambda$, $\delta\theta =\lambda$, $\delta
 H_{\mu\nu\rho}=0$, $\delta C =0$. We shall name parent Lagrangian invariance
 ``off-shell invariance'' meaning that $H$ field  has not been determined by its
 equation of motion. The  conpensator will disappear from the ``on-shell''
 Lagrangian obtained by solving the field equations for $H$. These equations are
 obtained by varying the $C$ field in (\ref{parent}), which gives 

\begin{equation}
{\partial L_P\over \partial C}=0\quad\longrightarrow\quad 
\partial_{[\,\lambda}\, H_{\mu\nu\rho\,]}=0 \qquad \Rightarrow \qquad
H_{\mu\nu\rho}= \partial_{[\,\mu}B_{\nu\rho\,]}
\label{bh}
\end{equation}

Eq.(\ref{bh}) gives Bianchi Identities for $H_{\mu\nu\rho}$ which has known
solution in terms of Kalb-Ramond field strength $H(B)$. Thus the Lagrange
multiplier $C$ is dualized to the tensor $B_{\mu\nu}$ following the same
pattern as in electric/magnetic duality case.
Inserting  solution (\ref{bh}) in the parent Lagrangian (\ref{parent}), 
one finds a $B\wedge F$ model described by the Lagrangian 

\begin{equation}
L_B =-{1\over 2\cdot 3!}\, H^{\, 2}{}_{\mu\nu\rho}(B)  -
{m\over 4}\, \epsilon^{\mu\nu\rho\sigma} \, B_{\mu\nu}\, 
\partial_{\,[\,\rho }\, V_{\sigma\,]}-
{1\over 4}\,  F_{\mu\nu}(\, V\, ) \, F^{\mu\nu}(\, V\,)
\label{lb4}
\end{equation}

On the other hand,
St\"uckelberg-like  Abelian model for $V_\lambda$, dual to the $B\wedge F$
model,  is obtained  varying (\ref{parent})  with respect to $H$ which gives
the equation of motion

\begin{equation}
H^{\mu\nu\rho}= \epsilon^{\mu\nu\rho\sigma} \,\left[\, \partial_\sigma C
-m\,\left(\, V_\sigma - \partial_\sigma\,\theta \,\right)\,\right]
\label{H}
\end{equation}

Re-inserting  (\ref{H})  in (\ref{parent})   one finds gauge invariant,
 massive Proca theory described by the St\"uckelberg   Lagrangian 

\begin{equation}
L_{\bf \Phi}= -{1\over 2}\left[\, m\,
\left(\, V_\mu - \,\partial_\mu\theta\right) - \partial_\mu C\,\right]^2-
{1\over 4}\, F_{\mu\nu}(\,V\, ) \, F^{\mu\nu}(\,V\,)
\label{fi4}
\end{equation}

The field $C$ is inert under gauge transformation and one can define a new
scalar field $\overline\theta = m\, \theta + C $, which is still a good
conpensator. Thus one obtains a familiar looking St\"uckelberg Lagrangian

\begin{equation}
L_{\bf \Phi}= -{1\over 2}\,
\left(\, m\, V_\mu - \,\partial_\mu\, \overline \theta\, \right)^2-
{1\over 4}\, F_{\mu\nu}(\,V\, ) \, F^{\mu\nu}(\,V\,)
\label{fi5}
\end{equation}

The legitimate question raises itself: is there a  simple generalization 
of the above procedure in case of a non-Abelian internal symmetry, 
e.g. $SU(N)$ ?\\
Earlier attempts has been made in this direction \cite{pkt} but the resulting
$B\wedge F$ Lagrangian was  found to have a non--linear kinetic term for the $B$
 field. The reason for this non-linearity can be traced to the problem of the
 {\it linear} kinetic term of the $B$  field with respect to  
the Kalb--Ramond  (KR) symmetry $\delta B_{\mu\nu}= \partial_{[\, \mu} 
\Lambda_{\nu\,]}$ in the non--Abelian case. To clarify the above comments,
let us look at the transformations of various fields,  which are

\begin{eqnarray}
&& {\bf V}_\mu{}^{\, \prime} ={\bf g}^{-1} \,{\bf  V}_\mu\,  {\bf g}
-i\,\left(\, \partial_\mu {\bf g}^{-1}\,\right){\bf g}\label{trasf1}\\ 
&&{\bf B} _{\mu\nu}{}^{\, \prime} ={\bf g}^{-1}\,{\bf B}_{\mu\nu}\, {\bf g}
\label{trasf2}
\end{eqnarray}

where, ${\bf g}$ stands for the group element of $SU(N)$.
The field strength of the non--Abelian Kalb--Ramond  field is defined as

\begin{equation}
 {\bf H}_{\mu\nu\rho}(\, {\bf B}\,)\equiv D_{[\,\mu }(\,{\bf V}\,)\, 
 {\bf B}_{\nu\rho\,]}=
 \partial_{[\,\mu }\, {\bf B}_{\nu\rho\,]} -i\,\left[\, {\bf V}_{[\, \mu},
 {\bf B}_{\nu\rho\, ]   }\,\right]
\end{equation}
 
 In addition to the usual $SU(N)$ Yang--Mills (YM) gauge transformations 
 one expects an  {\it additional } KR symmetry   described by the 
 vector parameter ${\bf \Lambda}_\mu $ as 
\footnote{We use the $\dagger$ and $\prime$ to denote vector and $SU(N)$
transformations in order to distinguish them. {\it N.B. $\dagger$ has nothing 
to do with hermiticity properties}.}

\begin{eqnarray}
&& {\bf V}_\mu {}^{\, \dagger} = {\bf V}_\mu \nonumber\\ 
&&{\bf B}_{\mu\nu}{}^{\, \dagger} ={\bf B}_{\mu\nu} + D_{[\,\mu }(\, {\bf V}\,)
\, {\bf \Lambda }_{\nu\,]}\label{kb}
\end{eqnarray}

The presence of the covariant derivative in the  transformation (\ref{kb})  
is induced by the fact that ${\bf B}_{\mu\nu}$ is a vector with respect to
$SU(N)$, thus the parameter ${ \bf \Lambda}_\mu$ also 
lives in the adjoint representation of $SU(N)$.  Transformations (\ref{kb})  
fails to be a symmetry of the non--Abelian version  of the kinetic term for
$B$ field. In fact, kinetic term transforms as 

\begin{equation}
 {\bf H}^{\, \dagger}{}_{\mu\nu\rho}(\,{\bf B}\,) =
 {\bf H}_{\mu\nu\rho}(\,{\bf B}\,) +i\, 
 \left[\, {\bf F }_{ [ \, \mu\nu}(\,{\bf V}\,), {\bf \Lambda} _{\rho \,] }\,
 \right] \label{comm}
 \end{equation}

and is clearly KR non--invariant. 
while $B\wedge F$ term remains invariant since its variation under (\ref{kb})
reduces to the non-Abelian Bianchi Identities of 
${\bf F}_{\mu\nu}(\, {\bf V}\,)$. One realizes that the non--Abelian extension
of the $B\wedge F$ model introduces a Kalb--Ramond gauge non--invariance already
at the level of kinetic terms, let alone mass terms. This problem has already
been noticed previously \cite{pkt} and by--passed avoiding a kinetic
term for  the non--Abelian  Kalb--Ramond gauge field in the parent Lagrangian.
The outcome of dualization was a non--linear $B$ interaction. 
We shall adopt a different approach and try to make an invariant kinetic term
with respect to both KR and YM symmetry which is linear. In this way we want 
to maintain similarity, as much as possible, to the  Abelian case previously 
described. In fact, a simple way to re-establish KR invariance of the kinetic
term in (\ref{comm}) is to require 

\begin{equation} 
\left[\, {\bf F }_{ [ \, \mu\nu}(\,{\bf V}\,), {\bf\Lambda} _{\rho \,]}\,\right]
=0
\end{equation}

This constraint has a  solution in terms of a  ``{\it flat 
connection}'' ${\bf V}_\mu= {\bf \Phi}_\mu\equiv {\bf U}^{-1}\partial_\mu
{\bf U}$ described in terms of the group element $U$. 
Thus, the moral of the story is that a  kinetic term for the $B$ field
{\it can} be made KR invariant  if the covariant
derivatives are defined in terms of the flat connection ${\bf \Phi}_\mu$
while maintaining its linear (Abelian--like) form.
Thus, appropriate KR symmetry transformations are 

\begin{eqnarray}
&& {\bf \Phi}_\mu {}^{\, \dagger} = {\bf \Phi}_\mu \nonumber\\ 
&&{\bf B}_{\mu\nu}{}^{\, \dagger} ={\bf B}_{\mu\nu} + 
D_{[\,\mu }(\, {\bf \Phi}\,)
\, {\bf \Lambda }_{\nu\,]}\label{kbflat}
\end{eqnarray}
 
At this point, we construct a non--Abelian version of the parent Lagrangian 
(\ref{parent})as  

\begin{eqnarray}
L_P=&-&{1\over 2\cdot 3!}\,Tr\, {\bf H} _{\mu\nu\rho}
{\bf H}^{\mu\nu\rho}
+{1\over 3!} \epsilon^{\mu\nu\rho\sigma} \,Tr\,{\bf C}\,
D_{[\,\mu}(\,{\bf \Phi}\,){\bf H}_{\nu\rho\sigma\,]} \nonumber\\
&+&{m\over 3!} \epsilon^{\lambda\mu\nu\rho}\, 
   Tr\, {\bf H}_{\lambda\mu\nu}\,\left(\,
 {\bf A}_\rho-{\bf \Phi}_\rho\,\right)    
-{1\over 4}\,Tr\, {\bf F}_{\mu\nu}(\, {\bf A}\,) {\bf F}^{\mu\nu}(\, {\bf A}\,)
  \label{parent1}
\end{eqnarray}

where, ${\bf \Phi}_\mu$ is the non--Abelian version for the Abelian conpensator 
$\partial_\mu \theta $ in (\ref{parent}), while the the other fields maintain
their role. The non--Abelian dualization follows the previously described
steps. Varying (\ref{parent1}) with respect to ${\bf C}$ gives the solution

\begin{equation}
{\partial L_P\over \partial {\bf C}}=0\quad\longrightarrow\quad 
D_{[\,\lambda}({\bf \Phi})\,{\bf H}_{\mu\nu\rho\,]}=0 \qquad 
\Rightarrow \qquad
{\bf H}_{\mu\nu\rho}=D_{[\,\mu}(\,{\bf \Phi}\,){\bf B}_{\nu\rho\,]} 
\label{solh}
\end{equation}

Solution (\ref{solh}) is the non-Abelian version of (\ref{bh}). One should 
notice
that this solution exists only if ${\bf \Phi}$ is flat, which points out the
importance of such choice. In the case of an arbitrary (non-flat) connection
a solution of the type (\ref{solh}) cannot be found due to the fact that
$\left[\,  D_{\mu}(\,{\bf V}\,)\ , D_{\,\nu}(\,{\bf V}\,)\,\right]=
{\bf F}_{\mu\nu}(\, {\bf V}\,)\ne 0$.\\ 
Inserting (\ref{solh}) into (\ref{parent1}) gives the non--Abelian
version of the $B\wedge F$ Lagrangian (\ref{lb4}) as

 \begin{eqnarray}
L_{\bf B}=&-&{1\over 2\cdot 3!}\,Tr\, {\bf H}_{\mu\nu\rho}(\,{\bf B}\ ,
{\bf \Phi}\,)
 \, {\bf H}^{\mu\nu\rho}(\,{\bf B}\ ,{\bf \Phi}\, )+
{ m\over 4 }\, \epsilon^{\mu\nu\rho\sigma} \,Tr\,{\bf B}_{\mu\nu} \,
D_{[\,\rho}(\,{\bf \Phi}\,)
 \left(\, {\bf A}- {\bf \Phi}\,\right)_{\sigma\,]}  \nonumber\\ 
&-&{1\over 4}\,Tr\, {\bf F}_{\mu\nu}(\,{\bf A}\,) \,{\bf F}^{\mu\nu}(\,{\bf 
A}\,)
\label{lb6}
\end{eqnarray}

 where, we have kept explicitly the combination
  ${\bf A}_\sigma -  {\bf \Phi}_\sigma$
 in the $B\wedge F$ term, following from the parent Lagrangian,
 in order  to make evident gauge invariance of this term,
 although $D_{[\rho}({\bf \Phi}){\bf \Phi}_{\sigma\, ]}\equiv 0$ for a flat
 connection.
\begin{eqnarray}
&& {\bf H}_{\mu\nu\rho}(\, {\bf B}\ , {\bf \Phi}\,)\equiv 
 D_{[\,\mu }(\,{\bf \Phi}\,)\, {\bf B}_{\nu\rho\,]}=
 \partial_{[\,\mu }\, {\bf B}_{\nu\rho\,]} -i\,\left[\, {\bf \Phi}_{[\, \mu}\, ,
\, {\bf B}_{\nu\rho\, ]}\,\right]\\
 &&{\bf F}_{\mu\nu}(\,{\bf A}\,)\equiv  D_{[\,\mu }(\,{\bf A}\,)\, {\bf
 A}_{\nu\,]}
\end{eqnarray}

On the other hand, Varying (\ref{parent1}) with respect
to ${\bf H}$ gives the field equation 

\begin{equation}
{\bf H}^{\mu\nu\rho}= \epsilon^{\mu\nu\rho\sigma} \,\left[\, 
D_\sigma({\bf\Phi} ){\bf C} -m\,\left(\,{\bf A }_\sigma - 
{\bf\Phi}_\sigma\,\right)\,\right]
\label{DAH}
\end{equation}

Inserting solution (\ref{DAH}) in (\ref{parent1})   reproduces the non--Abelian
version of the St\"uckelberg Lagrangian (\ref{fi5}) as

\begin{equation}
L= -{1\over 4}\,Tr\, {\bf F}_{\mu\nu}(\, {\bf A}\,) {\bf F}^{\mu\nu}(\, {\bf
  A}\,)-{1\over 2}\,Tr\,\left[\, m \, \left(\,  {\bf A}_\mu - {\bf \Phi}_\mu
  \, \right) -D_\mu({\bf \Phi}){\bf C}\,\right]^2
\label{l20}
\end{equation}

One can see from (\ref{lb6}) and (\ref{l20}) the ${\bf \Phi}_\mu$ plays
the role of {\it unique } conpensator both for the KR and YM symmetries
and does not take part in the dualization process.
On the other hand, it is the scalar field ${\bf C}$ which is dual
to ${\bf B}$, which is a standard $p$--duality correspondence.

As a conclusion, we have succeeded in formulating the non--Abelian  extension of
the dualization procedure described in \cite{top}. The key of the success can be
found in a  formulation of the covariant derivative with respect tot he flat
connection ${\bf \Phi}_\mu$, which is the simultaneous conpensator for KR and YM
symmetries. It is present in both versions of the theory and, thus, it is not
dualized.  


\end{document}